\input epsf
\documentclass[nohyper,notoc]{article} 
\usepackage{epsfig}

\def\beq{\begin{equation}}
\def\eeq{\end{equation}}
\def\bea{\begin{eqnarray}}
\def\eea{\end{eqnarray}}
\def\bq{\begin{quote}}
\def\eq{\end{quote}}

\def \gsim{\mathrel{\vcenter
     {\hbox{$>$}\nointerlineskip\hbox{$\sim$}}}}
\def\gappeq{\mathrel{\rlap {\raise.5ex\hbox{$>$}}
{\lower.5ex\hbox{$\sim$}}}}
\def\lappeq{\mathrel{\rlap{\raise.5ex\hbox{$<$}}
{\lower.5ex\hbox{$\sim$}}}}

\def\Rpv{R_p \! \! \! \! \! \! /~~}
\def\Lv{L \! \! \! /~}

\def\bea{\begin{eqnarray}}   
\def\eea{\end{eqnarray}}

\begin{document}
\vspace*{-1in}
\renewcommand{\thefootnote}{\fnsymbol{footnote}}
\begin{flushright}
~ \\
\texttt{hep-ph/yymmddd} 
\end{flushright}
\vskip 5pt
\begin{center}
{\Large {\bf Various definitions of Minimal
Flavour Violation for Leptons
}}
\vskip 25pt
 Sacha Davidson $^{1,}$\footnote{E-mail address:
s.davidson@ipnl.in2p3.fr} and  Federica Palorini
$^{1,}$\footnote{E-mail address:
f.palorini@ipnl.in2p3.fr
} 
 
\vskip 10pt  
$^1${\it IPNL, Universit\'e CB Lyon-1, 4 rue Enrico Fermi, 69622 Villeurbanne,France}\\
\vskip 20pt
{\bf Abstract}
\end{center}
\begin{quotation}
  {\noindent\small 
Neutrino masses imply the violation of lepton flavour 
and  new physics beyond the Standard Model.
However,  flavour change has only been observed in oscillations. 
In analogy with the quark sector, we could deduce  
the existence of a principle of Minimal Flavour Violation also for Leptons.
Such an  extension is not straightforward, since the 
mechanisms generating neutrino masses are unknown 
and many scenarios can be envisaged.
Thus, we explore some possible definitions of MFVL 
and propose a  notion that can include many models.
We build an R-parity violating neutrino mass model 
in agreement with our preferred definition of  MFVL,
 and show that flavour
violating processes  are not neccessarily  controlled by  the  MNS 
mixing matrix.

\vskip 10pt
\noindent
%PACS number(s):~12.60.Jv, 14.60.Pq, 11.30.Fs
}

\end{quotation}
%\noindent{Neutrino Physics,
%Supersymmetric Standard Model, Solar and Atmospheric Neutrinos}

\vskip 20pt  

\setcounter{footnote}{0}
\renewcommand{\thefootnote}{\arabic{footnote}}

%%%%%%%%%%%%%%%%%%%%%%%%%%%%%%%%%%%%%%%%%%%%%%%%%%%%%%%%%%%%%%%%%%%%%%
%% Introduction %%%%%%%%%%%%%%%%%%%%%%%%%%%%%%%%%%%%%%%%%%%%%%%%%%%%%%
%%%%%%%%%%%%%%%%%%%%%%%%%%%%%%%%%%%%%%%%%%%%%%%%%%%%%%%%%%%%%%%%%%%%%%
%\newpage
\section{Introduction}

Minimal flavour violation\cite{DGIS,Paolo} in the quark sector,  
is a useful framework
in which to construct TeV-scale models of New Physics.  It is predictive,
and includes many or most models that are consistent with quark
flavour data.  Recently, a definition of Minimal Flavour Violation (MFV)
has been introduced for leptons \cite{CGIW}. The proposed formulation
is predictive---it implies that
lepton flavour violation is determined  by
the light neutrino mass matrix--- but includes few of the many 
neutrino mass models \cite{numass,seesaw,RPV,Zee,triplet,anna}  
that are consistent with current observations.

The flavour-changing mixing angles
of the leptonic sector (MNS matrix),
are  not measured with the overconstrained precision
of the CKM matrix.  So MFV is not strongly suggested for leptons,
as it is  for quarks. However,
if one {\it assumes} that there is new physics at
the TeV-scale,  that satisfies MFV or a similar principle
in the quark sector, then it is reasonable to 
expect a similar principle to apply for leptons.
So it is interesting to explore different possible
definitions of minimal flavour violation for leptons (MFVL), 
and in particular
to study whether  it implies that lepton flavour violation is
 controlled by the
MNS matrix and the light neutrino masses.

In this paper,  we  take the principle of  MFV 
to limit the number of flavour structures allowed
to the renormalisable couplings of the theory. 
This flexible definition can be applied to
many models, but is less predictive than 
\cite{CGIW}. We explicitly construct an
R-parity violating neutrino mass model
that is ``minimally flavour violating'', 
in agreement with observation, and where
the lepton flavour violation is not controlled
by the light neutrino mass matrix.

In section \ref{review}, we review minimal flavour
violation for the quarks, and classify neutrino
mass generation mechanisms.  In section \ref{mfvl},
we discuss  the purpose of  Minimal Flavour Violation 
for leptons, and   various possible implementations
which we apply  to  some 
 neutrino mass models. In section
\ref{lambda}, we build an R-parity violating neutrino
mass model, using the $\lambda LLE^c$ coupling,
that satisfies our preferred definition of MFVL. 
In the Appendix is sketched a model satisfying
a more restrictive definition of MFVL.

\section{Review}
\label{review}

 Beyond-the-Standard-Model physics, in the form
of new particles or new interactions,  must
exist at some scale, to explain observations
such as dark matter, neutrino masses, the
baryon asymmetry and the temperature fluctuations
in the Cosmic Microwave Background. 
New physics at the TeV-scale (such as,
for instance, supersymmetry) is particularily  desirable because
it could be discovered at the LHC, and  would be theoretically welcome
to  address the hierarchy
problem. However, if there  are
new flavoured TeV-scale particles, as one would like,
it is puzzling that their footprints have not been
seen in rare flavoured and CP violating processes. 
So Minimal Flavour Violation  is introduced as a constraint on
the interactions of such  new particles, 
 to suppress their contributions
to flavoured observables.

We follow  the  approach to Minimal Flavour
Violation of \cite{DGIS}, which starts from the
flavour transformation properties of various
terms in the SM Lagrangian. We define the SM
to have massless neutrinos.
In three generations, the fermionic  kinetic terms
\beq 
\overline{q}_L D \!\!\!\! /~ q_L + 
\overline{u}_R D \!\!\!\! /~  u_R + 
\overline{d}_R D \!\!\!\! /~  d_R + 
\overline{\ell} D \!\!\!\! /~ \ell + 
\overline{e}_R D \!\!\!\! /~ e_R 
\eeq
have  a global $U_q(3)\times U_u(3)\times U_d(3)\times U_\ell(3)\times 
U_e(3)$  flavour symmetry.
For instance, $q_L$ is a three component vector
in  quark doublet flavour space, whose kinetic term is
invariant under $q_L \rightarrow V_q q_L$, where $V_q \in U_q(3)$.
This  large
symmetry group is broken to $ U_B(1)
\times U_{L_e}(1) \times U_{L_\mu}(1)
\times U_{L_\tau}(1) $  by the
Yukawa couplings
\beq
 \overline{q}_L {\bf Y_u } H_u u_R + 
 \overline{q}_L {\bf Y_d } H_u^c d_R + 
 \overline{\ell} {\bf Y_e } H_u^c e_R  + h.c.
\eeq
where $H_u$ is the SM Higgs, and  the index order on Yukawa matrices
is left-right.  In the lepton sector,
there is   one ``symmetry-breaking'' operator,
or ``spurion'' in the language of
\cite{DGIS},  per vector space: $ {\bf Y_eY_e }^\dagger$ in $\ell_L$ space, 
$ {\bf Y_e }^\dagger {\bf Y_e }$ in $e_R$ space. 
These hermitian matrices  can be diagonalised,  and
are uniquely identified by their eigenvalues in 
the eigenbasis.  So we
will sometimes say the operators can ``choose  a basis'',
and discuss interchangeably the matrix, the spurion
and the basis of eigenvectors who are normalised
to  have length$^2$ = the eigenvalue.
In the presence of ${\bf Y_e} $ (and the absence of other
``basis choosing'' operators in the lepton sector),
there are three remaining global $U(1)$s.  The
three conserved quantum numbers 
can be taken as the individual lepton flavours
\footnote{ The three $U(1)$s can also
be taken to correspond to the
three diagonal generators of $U(3)$ $ = \{I, \lambda_3, \lambda_8 \}$,
acting simultaneously
on the $\ell_L$ and $e_R$ flavour spaces.  
In this case one  conserves 
the total lepton number $L_e + L_\mu + L_\tau$, and the 
flavoured asymmetries  $L_e - L_\mu$ 
and  $L_e + L_\mu - 2 L_\tau$ }. So in our restricted
definition of the SM, neutrinos are massless and 
lepton flavours are conserved. We add neutrino masses
at the end of the section.

In the quark sector, $ {\bf Y_d } {\bf Y_d}^\dagger $
and  $ {\bf Y_d}^\dagger{\bf Y_d }$  choose
respectively  a basis in the $q_L$  and the $d_R$  
flavour spaces. Similarly, ${\bf Y_u} {\bf Y_u }^\dagger  $
and  $ {\bf  Y_u }^\dagger{\bf Y_u}$  choose
respectively  a basis in the $q_L$  and the $u_R$  
flavour spaces.  So  there are
two   operators in $q_L$ space,
 ${\bf Y_d} {\bf Y_d }^\dagger$ and  ${\bf Y_u} {\bf Y_u }^\dagger$, 
who  are not simultaneously diagonalisable.
Flavour is therefore not conserved,
and the misalignment between the two eigenbases is
parametrised by the CKM matrix.

 The mixing angles and phase of
the quark sector  are
over-determined in many 
   flavour-changing, flavour-conserving
and CP violating processes of the quark sector.
For instance, the  CKM angles  can be obtained  in tree level
processes, and used to predict  rates  that
are mediated by loops in the Standard Model. To date,
the experimentally measured rates agree with these predictions,
implying that  the  new physics contribution  in loops should be smaller
than the SM.
For new particles 
with generic flavour-changing couplings,
this  is a strong constraint,  placing the mass
above 10-100 TeV  \cite{CKM}.

 Minimal Flavour Violation was introduced  to allow New  Physics 
 to have TeV-scale masses, and be consistent
 with precision flavoured data from the quark sector.
 It is a restriction on
the flavour structure of new interactions.
The only
  operators allowed  in   the ``flavour-spaces'' are
 those of the SM (and the identity matrix). 
So flavour-change  and CP violation in the quarks are proportional to
the CKM matrix and quark Yukawa eigenvalues, eg
 to ${\bf Y_d Y_d}^\dagger$ in  the mass  basis of up-type
doublet quarks.  
MFV   is therefore  a predictive framework, 
and   encompasses many  of the  models that fit
the data.

Flavour-changing processes are also observed in the lepton
sector, in neutrino oscillations.  The weakly intereacting 
neutrinos are observed to have small mass differences,
and large mixing angles with respect to the charged leptons.
That is, in the lepton doublet space, there are two
operators that break the flavour $U_\ell(3)$ symmetry.
These are the charged lepton and neutrino mass matrices,
which ``choose bases'' related by the MNS matrix $U$.
In the charged lepton mass eigenstate basis (referred
to as the ``flavour'' basis), the light neutrino mass
matrix  satisfies
\beq
[m_\nu][m_\nu]^\dagger = U^* D_{m_\nu}^2 U^T 
\eeq
where $D^2_{m_\nu} = 
{\rm diag} \{ m_1^2, m_2^2, m_3^2 \}$.
To date,  only flavour changing charged current processes
(mediated by $W$ exchange) are observed in the lepton sector,
and MFV is not ``required'' for the leptons.
Four 
elements of the MNS matrix are measured---the remainder
being  obtained from  unitarity\cite{Belen}---and 
CP violation is not observed. This means new
leptonic physics  is not stringently
constrained to agree with SM predictions for
CP violation, as is the case in the quark sector. 
Rates for  unobserved  FCNC lepton processes 
({\it e.g.}   $\mu \rightarrow e \gamma$)
can be calculated using the MNS matrix and neutrino masses,
and  are well
below the current experimental sensitivity. 
So new leptonic physics is only constrained 
to be less than the experimental rates, and not,
as in the quark sector, to be smaller  than
the prediction one obtains using observed masses.

The neutrino masses can be lepton number conserving
(``Dirac'') or not (``Majorana'').
In the Dirac case one could define MFV  in the lepton
sector as an exact copy of the quarks, so
in this paper, we consider  Majorana neutrino masses,
which  arise from   a dimension five operator
\beq
 (\ell_j H_u) {\bf K}^{jk} (\ell_k H_u) ~~.
\label{K}
\eeq
Two classes of new physics
 generating this operator can be distinguished.   One possibility
is that it is generated by new flavoured particles,
in a new flavour space. These new particles
should be heavy or weakly coupled, since they
have not been observed. The canonical example 
is the seesaw, where one adds, {\it e.g.}, 3 generations
of $\nu_R$, and the flavour symmetry group of
the kinetic terms is enlarged to $U(3)^6$. 
The second possibility is that the all
flavoured particles live in the 5 flavour spaces of the SM,
and some new lepton number- or flavour-changing
interactions are included. 
This is the case for neutrino masses generated
in the R-parity violating MSSM.

\section{Minimal Flavour Violation for Leptons?}
\label{mfvl}

We   {\it assume} that there are new flavoured particles
at the TeV scale, and  hope that  this is verified
soon at the LHC.
A definition of Minimal Flavour Violation in the
lepton sector \cite{CGIW} could  then be interesting 
for various reasons. Firstly, MFV in the quark
sector is well motivated by the experimental
observations.  So one could conclude it reflects
some principle or symmetry of the underlying theory,
and  should apply in the lepton sector as well.
 Secondly, in the lepton sector, we know there must be
Beyond the Standard Model physics
at some scale, because we observe neutrino masses. 
We can hope  to use  MFV as  a tool
in distinguishing among the multitude of
candidate models for new physics in the lepton sector
\footnote{Taking a 
principle of MFV to apply to the neutrino mass generation
mechanism is a more ambitious implementation of
MFV than in the quark sector. 
For quarks, one hopes for new TeV-scale particles 
(for instance to address the hierarchy problem), 
in which case  MFV is almost required to
describe the Lagrangian up to scales $\sim 100 $ TeV.
In the lepton sector,  we know there is New Physics,
and it should have some connection to flavour because
it generates neutrino masses.
However, this new physics could
be at a high scale ($\sim 10^{10} - 10^{16}$ GeV in the seesaw?),
so in assuming that MFV applies to the
interactions that generate the neutrino masses,
we may be applying it across many
more orders of magnitude than in the quark sector.}.
Minimal Flavour Violation for leptons should therefore
be applicable to  most models, and be predictive, so we can  
test the hypothesis and/or differentiate models.

A predictive definition of MFV for  the lepton sector has recently
been introduced in \cite{CGIW}, and further  studied in \cite{CG}.
It supposes that
the three light neutrinos are  Majorana, with
the required lepton number violation occuring
at some high scale $\Lambda_{LN}$.
 Two classes of models 
are  considered: those  whose particles  transform according
to the $U^5(3)$ flavour transformations of the SM,
and  a second scenarino with three (heavy)
right-handed neutrinos. 

In this work, we also take  
the light neutrino masses  to be Majorana.
If they were Dirac, MFV could be defined for leptons by copying the
quark definition. Models that generate Majorana neutrino
masses can be divided into two cases \cite{CGIW}: \\
$\bullet$ case A:    models whose particles  transform according
to the $U^5(3)$ flavour transformations of the SM, and \\
$\bullet$ case B:  models with  a
flavour transformation  group  that is
larger than that of the SM ({\it e.g.} the seesaw, where the
  kinetic terms of the three $\nu_R$ have a $U(3)$ symmetry).

\subsection{Larger flavour transformation group}

Suppose there are a several generations of a
new particle,  {\it e.g.}  three right-handed neutrinos.
The kinetic terms therefore have an enlarged flavour
symmetry group, which is $U^6(3)$ when  3 $\nu_R$
are added to the SM.  
The renormalisable Lagrangian for the SM + the
new particle will contain the  SM Yukawas,
and some number of additional spurions  corresponding to
the interactions of the new particle. In the case of
the seesaw, the Lagrangian is
\beq
 {\cal L}_{SM} +  \overline{\ell}_j {\bf Y_\nu }^{jK} H_u N_K + 
 \frac{1}{2} \overline{N_J^c} {\bf M}_{JJ} N_J + h.c. 
\label{seesaaw}
\eeq
and there  are potentially two ``basis-choosing'' interactions,
or spurions, in $\nu_R$ space: ${\bf Y_\nu}^\dagger {\bf Y_\nu}$
and ${\bf M}$. There are a variety of potential 
definitions of MFV, which we illustrate with the
seesaw example.

\begin{enumerate}
\item  one could impose  that the new physics 
may not introduce new spurions in the Standard Model
flavour spaces. In the case of the seesaw, this means
that ${\bf Y_\nu} {\bf Y_\nu}^\dagger$ should be
diagonal in the  lepton doublet flavour basis 
(charged lepton mass eigenstate  basis). No restrictions
are imposed on the number of  bases chosen 
in  the flavour space of the new particles.
In the seesaw case,  ${\bf Y_\nu}^\dagger {\bf Y_\nu}$
and ${\bf M}$  could have different eigenbases, and must do so
to reproduce the correct neutrino mixing angles.
This definition of MFV for leptons is predictive but
unattractive, because it implies that lepton flavour
violation amoung charged leptons is suppressed by
neutrino masses.
\item   CGIW \cite{CGIW} define MFV for the seesaw by allowing
the renormalisable interactions of eqn (\ref{seesaaw})
to choose a second basis in $\ell$ space, but impose
restrictions on the spurions  in $\nu_R$ space. They  study  the case
where  the $\nu_R$ are degenerate of mass $M$, 
and CP is a symmetry of the right-handed neutrino sector
\cite{9810328}. 
(So there is only one eigenbasis in $\nu_R$ space.)

In this case $ {\bf K} = {\bf Y_\nu M^{-1} Y_\nu^T} =  
 {\bf Y_\nu  Y_\nu}^T/M$. 
The  two ``basis-choosing'' coupling matrices in $\ell$ space,
that are relevant for lepton number conserving 
flavour violation, 
are 
\beq
{\bf Y_e} {\bf Y_e}^\dagger ~~,~~
{\bf Y_\nu} {\bf Y_\nu}^\dagger = \frac{M}{v^2} {\bf U}^*{\bf D}_{m_\nu} 
{\bf U}^T
\label{diff}
\eeq
If there is no CP violation in the lepton sector
(the case studied by CGIW), then ${\bf Y_\nu} {\bf Y_\nu}^\dagger
= M {\bf K}$. In either case,  lepton flavour violation
is controlled by parameters from the light neutrino sector.  
The predictions of this  scenario should  be similar to
the SUSY seesaw with degenerate $\nu_R$ \cite{hisano1}.
\item  The more generic (and less predictive)
definition of MFV for the seesaw would be to
allow  all renormalisable interactions to
be independent spurions, as one
allows for SM constituents (equivalently, one
could allow up to  two spurions per vector space). In the
case of the $\nu_R$,   with the seesaw Lagrangian
of eqn (\ref{seesaaw}), 
the  Majorana mass matrix ${\bf M}$ and
the Yukawa coupling ${\bf Y_\nu } ^\dagger {\bf Y_\nu } $
are independent spurions in the $\nu_R$ flavour space.
Similarly to   ${\bf Y_\nu } {\bf Y_\nu } ^\dagger $
 and  ${\bf Y_e } {\bf Y_e } ^\dagger $ in $\ell$
space, they  have unrelated eigenbases. 
 This  is the ``usual'' type-1 seesaw, whose 
supersymmetric flavour-changing  predictions
have been extensively studied in the literature
\cite{LFVss}.   It is
well known that in the SUSY seesaw,  
the rates for  flavour-changing
processes among the charged leptons are not related
to the neutrino masses or the MNS matrix \cite{DI}.
\end{enumerate}

\subsection{Standard Model flavour transformations}

Consider now neutrino mass models 
 whose particles  transform according
to the $U^5(3)$ flavour transformations of the SM.
In the quark sector, MFV restricts the bases chosen by
 flavour-dependent new interactions to
be those of the SM Yukawas. That is, there are two
allowed bases  (spurions) in $q_L$ space, and
one in $u_R$ and $d_R$ spaces respectively.

\begin {enumerate}
\item   CGIW  define MFV  for leptons, in this case, to
 allow two spurions  (``basis-choosing'' operators)
in  the doublet lepton ($\ell$) space, which are ${\bf K}$
and  ${\bf Y_e} {\bf Y_e}^\dagger$ .
The  ${\bf K}$ is the dimensionful coefficient of a lepton number violating
operator, so lepton flavour changing processes, 
that conserve lepton number, are controlled by
the dimensionless $\Lambda_{LN}^2 {\bf K K}^\dagger$.
Rates for lepton flavour violating processes ({\it e.g.}
$\mu \rightarrow e \gamma$) are proportional to the
unknown $\Lambda^2_{LN}$, but ratios of LFV processes
are predicted to be controlled by 
${\bf K K}^\dagger$.
This describes for instance the SUSY triplet model \cite{anna,CGIW,CG}.
\item Alternatively one could
 suppose that MFV is a restriction on {\it renormalisable}
couplings.  This is reasonable firstly  because 
MFV  is a recipe for extrapolating in scale. We know
how renormalisable   couplings evolve, whereas 
we cannot guess, in a bottom up approach, when a
non-renormalisable interaction becomes
renormalisable. Secondly, one could expect
that flavour is  introduced into
the theory at some high scale, ($M_{GUT}$?), and comes to
us via renormalisable couplings.\\
$\bullet$ One could hope to define
MFV,  by analogy with the quark sector, 
as restricting all new interactions to
be aligned with the SM Yukawas. 
But then it is difficult to
obtain the large mixing angles of the  MNS matrix.
A model attempting to satisfy this ideal
can be found in Appendix A.
This version of MFV would 
predict that lepton flavour changing
amplitudes must contain the neutrino mass to some power,
or  lepton number violation. Notice that
this differs from the CGIW prediction;  in the present
case, 
lepton flavour violation is suppressed by the small
neutrino mass scale.\\
$\bullet$ A more realistic definition of MFV, that
includes some models, would allow
(at least) one other basis 
in $\ell$ space.
New renormalisable  interactions can
choose one, and only one,  new basis for $\ell$ space, and no
new bases for $\{e_R, u_R,d_R,q_L \}$  spaces. 
That is, we take MFV as a statement about
renormalisable interactions, that  allows two
bases in  the $q_L$ and $\ell$ spaces, one in
the $u_R, d_R$ and $e_R$ spaces.
The question then arises: 
is lepton flavour violation among charged leptons
controlled by the light neutrino mass matrix?
If yes, then this definition of MFVL is equivalent to
that of CGIW. If the lepton flavour violating rates
are independent, one could hope
they give information about the neutrino mass generation
mechanism.

Some   renormalisable, lepton number violating 
interactions involving $\ell$, that can be used
to construct the neutrino mass matrix, are
\bea
\frac{1}{2} M_T \vec{T} \cdot \vec{T}^\dagger + 
g_\phi  M_T H_u^c \vec{\tau} H_u^c  \cdot \vec{T}
+ g_{ij} \overline{\ell}_i^c \vec{\tau}\ell_j \cdot \vec{T} 
 &&{\rm (triplet)} \label{triplet} \\
\mu_i L_i H_u + \lambda'_{jrs} L_jQ_rD_s^c + 
\frac{1}{2}\lambda^n_{ij} L_iL_jE_n^c &&  {\rm (R~ parity~ Violating)}\label{RPV} 
\eea
where  $H_u$ is the  Standard Model  doublet Higgs,
$T$ is an SU(2) triplet scalar, and
 the second line is in superfield notation,
so are  renormalisable interactions 
in supersymmetry.  Under the $U(3)$ flavour transformations
of $\ell$ space,  $g$ transforms as a symmetric
${\bf 6}$, $\lambda'$ and $\mu$  as  
${\bf \bar{3} }$, and  the antisymmetric $\lambda$ as a
$ {\bf 3}$.

In the triplet model of
eqn (\ref{triplet}), the exchange of
$\vec{T}$  induces the neutrino mass operator  (\ref{K}).
 The light neutrino mass matrix is
therefore $[m_\nu]_{\alpha \beta} 
\propto g_{\alpha \beta}$,
and flavour violation among
the charged leptons is controlled by 
the light neutrino mass matrix \cite{anna}.   In this model,
this definition of MFV  based on renormalisable
couplings, agrees with the definition of CGIW
based on mass matrices.

It seems not possible to generate observed light
neutrino masses with the $\lambda'$ coupling, if we
implement strictly  this  definition of MFV.
The $\lambda'$ must respect MFV in the quark sector:
$$\sum_{\ell,d}  \lambda'_{\ell q d}\lambda^{'*}_{\ell p d}
\propto [{\bf Y_d}{\bf Y_d}^\dagger]_{qp} ~~~~~
\sum_{\ell,q}  \lambda'_{\ell q d}\lambda^{'*}_{\ell q f}
\propto  [{\bf Y_d}^\dagger{\bf Y_d}]_{df}$$
so the eigenvalues of $\lambda'$ are those of the ${\bf Y_d}$.
This hierarchy, when combined with quark masses to obtain
$m_\nu$, gives too steep a neutrino mass hierarchy.
In the following section, we construct a neutrino mass
model that satisfies this definition of MFV, using the
$\lambda$ interaction.
\end{enumerate}

\section{The $\lambda$ model}
\label{lambda}

The aim of this section is to construct a 
neutrino mass model that  has two features.
It should be minimally flavour violating,
in the sense that the new renormalisable interaction
$\lambda$ only introduces one new basis, or spurion, 
which is in $\ell_L$ space. And the model
should   agrees with current bounds on lepton flavour violating
processes ($\mu \rightarrow e \gamma$,  etc),
{\it but} the predictions for these processes
should not be determined by the light neutrino
mass matrix.

We take the light neutrino masses to be
generated entirely by the RPV $\lambda$ coupling, so we
neglect $\lambda'$ and bilinear RPV.  In the
charged lepton mass eigenstate basis, the light neutrino
mass matrix can be written \cite{DL2}
\beq
[m_\nu]_{ij}  =   \sum_{ m,n,p,q} 
\lambda_{in}^{m} \lambda_{pj}^{q} 
m_{e_{n}} \delta_q^n \tilde{A}^{mp}    
I( m_{\tilde{E}_{m}}, m_{\tilde{L}_{m}})  + (i \leftrightarrow j)
~~~, \label{mass}
\eeq
where  the $A$-term  $
\tilde{{\bf A}}^{mp} = -  
( ({\bf Y_e A})^{mp} \frac{v_d}{\sqrt 2} + 
\mu  \frac{v_u}{\sqrt 2} {\bf Y_e}^{mp} )
$  is taken flavour diagonal and included  in the mass
insertion approximation, 
the mass matrices for the  sleptons $\tilde{E}_m$ and
$\tilde{L}_m$  are  taken diagonal in the
flavour basis (which is consistent with MFV),
and 
\beq
 I(m_{1}, m_{2 }) = - {1\over 16\pi^2} 
{m_1^2\over m_1^2 -m_2^2} \ln\frac{m_1^2}{m_2^2} ~~.
\eeq

The $\lambda_{ij}^n$ is an antisymmetric matrix
on its doublet indices $i,j$, so  corresponds to a plane
in $\ell_L$ space. It is convenient to rewrite it
as a single index object in $\ell_L$ space (the
vector orthogonal to the plane), using the antisymmetric
$\epsilon$ tensor
\beq
\tilde{\lambda}_{nk} = \frac{1}{2} \epsilon_{ijk}\lambda_{ij}^n.
\label{epsijk}
\eeq 
The $\epsilon_{ijk}$ is SU(3) invariant, but not $U(3)$ invariant,
so this renaming has some peculiar consequences. 
Consider the case where $\lambda_{ij}^n \propto \epsilon_{ijn}$,
so $\tilde{\lambda}$ is ``flavour diagonal''. 
However, since it transforms under $SU_\ell(3) 
\times SU_e(3) $ as  $\ell e_R$, it is not invariant, 
and   the flavour differences  $L_e - L_\mu$ and 
$L_e + L_\mu - 2 L_\tau$ are conserved mod 2
in four fermion interactions.

The ``MFV'' constraint is that 
$\sum_j \tilde{\lambda}_{sj} \tilde{\lambda}_{tj}^{*}$ should
be diagonal in the singlet charged lepton mass basis, with
eigenvalues proportional to the charged lepton Yukawas \footnote{Since
the SM has  only one eigenbasis  in $e_R$ space,
it is  not required of new interactions that
they have the same eigenvalues as ${\bf Y_e }^\dagger {\bf Y_e}$,
provided that they have the same eigenvectors.}.
We will permute the $\mu-\tau$ eigenvalues, that is
$\sum _{ij} \lambda^e_{ij}   \lambda^{e*}_{ij} \propto m_e^2/v^2$,
but   $\sum _{ij} \lambda^\mu_{ij}   \lambda^{\mu*}_{ij} \propto m_\tau^2/v^2$,
and $\sum _{ij} \lambda^\tau_{ij}   \lambda^{\tau*}_{ij} \propto m_\mu^2/v^2$.
On its doublet indices
\beq
V_\lambda^\dagger \tilde{\lambda} \tilde{\lambda}^{\dagger} 
V_\lambda = {\rm diag } \{ m_e^2, m_\tau^2, m_\mu^2 \}/v^2 
\eeq
where $V_\lambda$ is a unitary matrix transforming
from the charged lepton basis to the eigenbasis
of $\tilde{\lambda}$. The observed
light neutrino parameters, with masses in the
inverse hierarchy, can be obtained from 
\beq
V_\lambda^\dagger = 
\left[ \begin{array}{ccc}
-c \epsilon &\frac{1+s\epsilon}{\sqrt{2}} &-\frac{1-s\epsilon}{\sqrt{2}}  \\
s&\frac{c}{\sqrt{2}}  &\frac{c}{\sqrt{2}}  \\
c&- \frac{s-\epsilon}{\sqrt{2}}  &-\frac{s+\epsilon}{\sqrt{2}}  
\end{array}
\right]
\label{V}
\eeq
where $c = \cos (\pi/4+\delta)$. 
This  corresponds to
\bea
\lambda^e_{\alpha \beta} \propto \frac{m_e}{v} \sim 0 \nonumber \\
\lambda^\mu_{\mu \tau} = s \frac{m_\tau}{v} && \lambda^\mu_{e \tau} = \frac{c}{\sqrt{2}}
 \frac{m_\tau}{v}  ~~~~  \lambda^\mu_{e \mu} = \frac{c}{\sqrt{2}}
 \frac{m_\tau}{v} \nonumber \\ 
\lambda^\tau_{\mu \tau} = c \frac{m_\mu}{v} && \lambda^\tau_{e \tau} =- \frac{s-\epsilon}{\sqrt{2}}
\frac{m_\mu}{v} ~~~~   \lambda^\tau_{e \mu} = -\frac{s+\epsilon}{\sqrt{2}}
 \frac{m_\mu}{v} \label{weuse}  
\eea

For  $\theta = \pi/4$, $\epsilon = 0$ and
degenerate sleptons, eqn (\ref{mass})  gives exactly degenerate  neutrinos
$\nu_e$ and $\nu_{\mu - \tau}$, whose mass varies inversely with
the slepton mass. 
The observed neutrino mass differences and mixing angles, in the inverse hierarchy,
can be obtained by including small perturbations.
e difference  between the square of the slepton masses $\tilde{m}_{\mu}^2 - 
\tilde{m}_{\tau}^2$ 
and the small mixing angle $\epsilon$  contribute to splitting
the $\nu_1$ and $\nu_2$ masses, 
 while the  
parameter 
$\delta$ of the $V_{\lambda}$ matrix seems to control  the solar mixing angle.

%%%%%%%%%%%%%%%%%%%%%%%%%%%%%%%%%%%%%
%%%%%%%%%%%%%%%%%%%%%%%%%%%%%%%%%%%%%
\begin{figure}[t!]

\centerline{\hspace{-3.3mm}
\epsfxsize=8cm\epsfbox{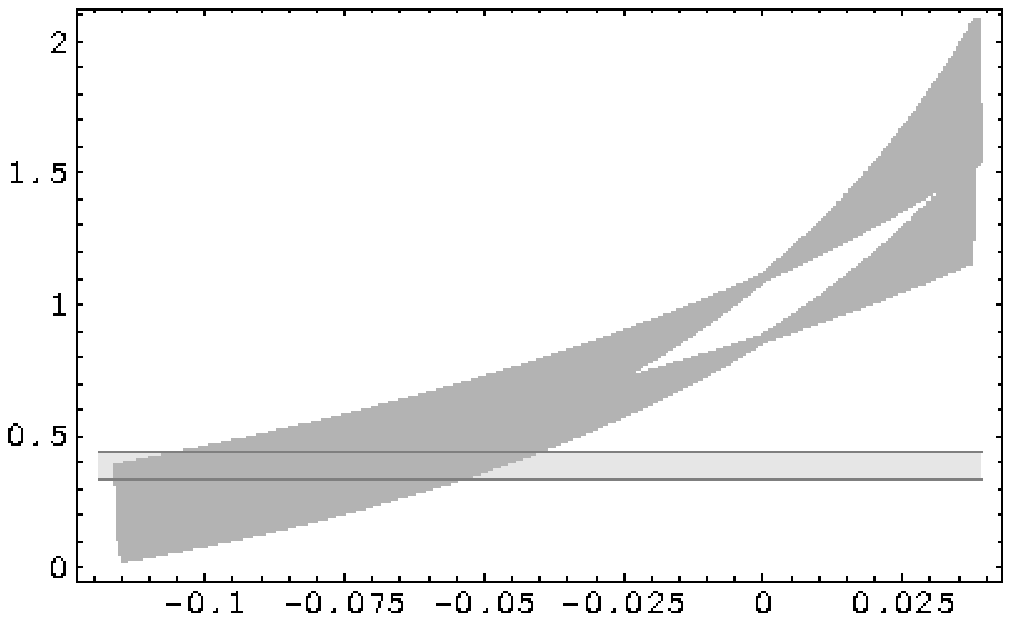}
\hspace{0.5cm}
\epsfxsize=8cm\epsfbox{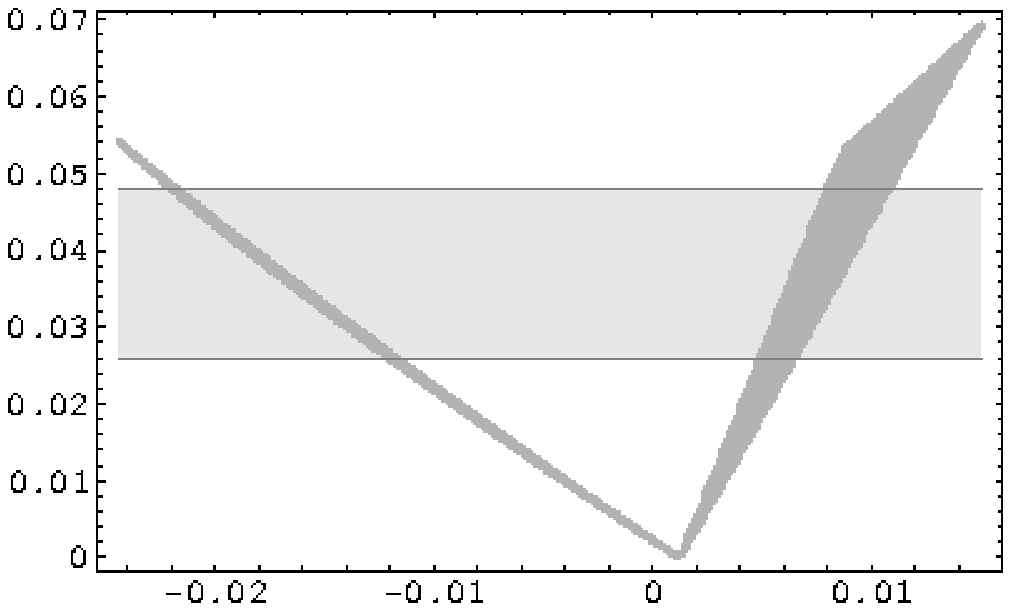}}

\caption{For slepton masses at $\sim 300$ GeV and $\tan\beta \sim 25$, 
on the left we plot our prediction for the ``solar'' mixing  parameter 
$\tan^2 \theta_{12}$ 
as a function of the model parameter $\delta$
(see eqn (\ref{V})); where we have choosen an $\epsilon$ 
such that the other parameters ($\Delta m_{sol}^2/\Delta m_{atm}^2$, 
$\tan^2\theta_{23}$, $\sin^2\theta_{13}$) satisfy the experimental bounds. 
On the right we
plot the ratio $\Delta m_{sol}^2/\Delta m_{atm}^2$ as a function 
of $\epsilon$, with $\delta$ consistent with the experimental bounds 
on the other physical parameters 
($\tan^2\theta_{12}$, $\tan^2\theta_{23}$, $\sin^2\theta_{13}$).
In both cases the horizontal light gray band  represents
the experimentally allowed parameter space
 \cite{neudata}.}

\protect\label{physparam}
\end{figure}
%%%%%%%%%%%%%%%%%%%%%%%%%%%%%%%%%%%%%%
%%%%%%%%%%%%%%%%%%%%%%%%%%%%%%%%%%%%%%

In Figure~(\ref{physparam}) we show the behaviour of two physical parameters, $\tan^2\theta_{12}$ and the ratio $\Delta m_{sol}^2/\Delta m_{atm}^2$, when the parameters $\delta$ and $\epsilon$ vary. For both the plots, we have considered only those points in agreement with the experimental bounds on the remaining set of neutrino parameters. From the  intersection between the dark and the light region we can deduce the range of availability for $\epsilon$ and $\delta$. (The slepton mass difference in these plots is fixed at a value that
could be generated by renormalisation group running.)

This $\lambda$ model, then, satisfies our requests. The neutrino masses are generated by a renormalisable operator $\tilde{\lambda}$, whose eigenbasis is related with the charged lepton mass eigenbasis by a matrix $V_{\lambda}^{\dagger}$ different from the MNS mixing matrix.
The matrix $V_{\lambda}^{\dagger}$, then, has become the operator that guides flavour violating processes, whose amplitudes are now determined by the $\lambda$ couplings. In particular, we can see in (\ref{weuse}) that the order of magnitude of each $\lambda$ coupling is determined by its upper index, which is related with the flavour of the right-handed particle involved in the vertex. As we can see in Table~\ref{tabFV} the experimental bounds on FV decays are satisfied in this $\lambda$ model, in agreement with our definition of MFV. 

The strongest experimental constraints on FV processes are given for the 
muon decay into three electrons and the 
$\mu \rightarrow e \gamma$ decay \cite{PDG,meg,meee}. 
The flavour violating decays with charged leptons in the initial 
and final states, like $\mu^- \rightarrow e^- e^+ e^-$, appear at the 
tree level and are mediated by the exchange of a left-handed sneutrino $\tilde{\nu}_i$. 
So, it can be easily understood why the decay rates of muon, but also tau, 
into three electrons are so low. In addition to the suppression due to the 
sneutrino mass, in each diagram appears a vertex $\tilde{\nu}ee$, whose 
amplitude is determined by a coupling of the form $\lambda_{\alpha \beta}^{e}$ 
which is proportional to the small electron mass, since right-handed sneutrinos 
are not present in the model.

The $\mu \rightarrow e \gamma$ decay \cite{meg}, instead, appears at a loop level, 
mediated by a charged lepton and slepton. In this case the main contribution 
comes from the diagram with vertices proportional to $\lambda_{\mu\tau}^{\mu} 
\lambda_{e\tau}^{\mu} \propto m_{\tau}^2/v^2$, whose large contribution is 
somewhat compensated by the loop suppression. We estimate  the decay branching 
ratio in our model to be  $\sim 6 \times 10^{-13}$, and we can notice that, 
although this value respects the present experimental constraint, it could be 
accessible in the next  experiments.

%\begin{center}
\begin{table}[t!]
\begin{center}
\begin{tabular}{|l|c|c|}
\hline
&&\\
&  \textbf{Expected value} & \textbf{Experimental bound}\\
&&\\
\hline
\hline
&&\\
$BR(\mu^- \rightarrow e^- e^+ e^-)$ & $\sim 10^{-17} \left( \frac{100 \ GeV}{m_{\tilde{\nu}}} \right)^4$&$< 1.0 \ 10^{-12}$ \\
$BR(\mu^- \rightarrow e^- \gamma)$&$\sim 10^{-12} \left( \frac{100 \ GeV}{m_{\tilde{\nu}}} \right)^4$&$< 1.2 \ 10^{-11}$\\
$BR(\tau^- \rightarrow e^- e^+ e^-)$ & $\sim 10^{-19} \left( \frac{100 \ GeV}{m_{\tilde{\nu}}} \right)^4$&$ < 2.9 \ 10^{-6}$ \\
$BR(\tau^- \rightarrow e^- \mu^+ \mu^-)$ & $\sim 10^{-10} \left( \frac{100 \ GeV}{m_{\tilde{\nu}}} \right)^4$&$< 1.8 \ 10^{-6}$ \\
$BR(\tau^- \rightarrow e^+ \mu^- \mu^-)$ & $\sim 10^{-12} \left( \frac{100 \ GeV}{m_{\tilde{\nu}}} \right)^4$&$< 1.5 \ 10^{-6}$ \\
$BR(\tau^- \rightarrow \mu^- e^+ e^-)$ & $\sim 10^{-12} \left( \frac{100 \ GeV}{m_{\tilde{\nu}}} \right)^4$&$< 1.7 \ 10^{-6}$ \\
$BR(\tau^- \rightarrow \mu^+ e^- e^-)$ & $0$&$< 1.5 \ 10^{-6}$ \\
$BR(\tau^- \rightarrow \mu^- \mu^+ \mu^-)$ & $\sim 10^{-11} \left( \frac{100 \ GeV}{m_{\tilde{\nu}}} \right)^4$&$< 1.9 \ 10^{-6}$ \\
&&\\
\hline
\end{tabular}\caption{Table of the branching ratios for flavour violating processes. In the second column appear the branching ratios predicted in the $\lambda$ model, while in the third column are indicate the experimental bounds at $90 \%$ of confidence level \cite{PDG}.}\label{tabFV}
\end{center}
\end{table}
%\end{center}

%%%%%%%%%%%%%%%%%%%%%%%%%%%%%%%%

\section{Summary}
In the lepton sector, new beyond-the-Standard-Model 
(BSM) interactions  are required
to generate neutrino masses. If these masses are
Majorana, they arise from a dimension five operator
whose flavour structure (eigenvalues, eigenvectors
in lepton doublet space) may not be the same as the 
(renormalisable) BSM  interactions.
In this context, it is not obvious to define
Minimal Flavour Violation for leptons. One can
take the ``minimal'' scenario to be that
flavour-change in the lepton sector is controlled
by the neutrino mass matrix, which is in part known.
This predictive approach was taken  in \cite{CGIW}.

However, there are neutrino mass generation mechanisms
that do not make this prediction, since flavour change may
be proportional to a different combination of
renormalisable couplings than enters $m_\nu$. In this paper,
we  explore various possible definitions of ``minimally
flavour violating'', based on the renormalisable
interactions in the Lagrangian.  We suppose that
Minimal Flavour Violation is a restriction on the
number of  inequivalent eigenbases that 
{\it renormalisable}  flavour-dependent
interactions can choose. 
The most minimal possibility would be to restrict the
new interactions to align themselves with the charged lepton Yukawa,
but then it is difficult to obtain MNS mixing angles
(without enlarging the flavour transformation
group, for instance by adding right-handed neutrinos).
The leptonic masses and mixing angles can be obtained in more models
({\it e.g.} triplet, R-parity violation) by allowing
two eigenbases in doublet lepton space. The second basis
may be other than the neutrino mass basis;  we  
construct  a   model where flavour violation
among charged leptons is not predictable from 
the light neutrino mass matrix.

Data in the quark sector suggest that new particles and
interactions at the TeV-scale should satisfy  Minimal
Flavour Violation. Data in the lepton sector do not
 require a minimal flavour violation principle,
but one could imagine it is there, by analogy with
the quarks. 
Unfortunately,  there are many possible definitions,  
which seem either predictive, or
able to include many
models.  A compelling
definition of MFV,   giving 
different predictions for   different
neutrino mass generation mechanisms,
could  be useful in attempting a bottom-up
reconstruction of the neutrino mass  mechanism
\cite{EFT}
from  lepton flavour violating rates.

\section{Appendix A}

The aim of this appendix is to obtain an acceptable
neutrino mass matrix, using new interactions that
are diagonal in  the charged lepton mass basis.
We consider an RPV model,  with lepton number
violating terms in the superpotential
\beq
\frac{1}{2} \lambda_{ij}^k L_iL_jE^c_k + \mu_iH_u L_i ~
\eeq
an $\Lv$ soft term $ B_i H_u L_i$, and
we estimate the light neutrino mass matrix from
the formulae in \cite{Davidson:2000uc}.
It seems possible to obtain degenerate light neutrinos
($m_\nu \sim $  0.2 eV), and an  MNS matrix in agreement
with observations \footnote{we did not scan parameter
space, more realistic masses could be possible. Our example
requires delicate fine-tuning.}.

This peculiar result arises by taking
\beq
\lambda^k_{ij} = \lambda \epsilon_{ijk} ~~,~~
\mu_\mu =  \delta_\mu  \mu_0~~, ~~  B_\tau =  \delta_B B_0
\eeq
and all other $\Rpv$  couplings to be zero.
The usual $\mu_0$ and $B_0$ terms are
$\mu_0 H_u H_d$ in the superpotential and
$B_0  H_u H_d$  among the soft breaking terms,
and we will later solve for the desired values of
$\delta_\mu, \delta_B$.
We claim that $\lambda \propto \epsilon_{ijk}$
is ``flavour-diagonal'', insofar as  it is an
SU(3) invariant (see discussion after eqn
(\ref{epsijk})). We can obtain off-diagonal
contributions to the neutrino mass matrix, by combining it
with the ``bilinear'' $\Rpv$  interactions $B_\tau$
and $\mu_\mu$.

The leading contributions to the neutrino mass matrix,
in the charged lepton mass basis, can be
estimated as \cite{Davidson:2000uc}
\beq
[m_\nu] \simeq
\frac{1}{8 \pi^2 m_{SUSY}}
\left[
\begin{array}{ccc}
 \lambda^2 m_\mu m_\tau &
\lambda \delta_B m_\tau  (h_\mu m_\mu - h_e m_e) &
\lambda \delta_\mu m_\mu  (h_e m_e - h_\tau m_\tau) \\
\lambda \delta_B m_\tau  (h_\mu m_\mu - h_e m_e) &
8 \pi^2 |\delta_\mu|^2 m^2_{SUSY} &
g^2 \delta_\mu \delta_B m^2_{SUSY}/8 \\
\lambda \delta_\mu m_\mu (h_e m_e - h_\tau m_\tau)
&g^2 \delta_\mu \delta_B m^2_{SUSY}/8&
g^2  \delta^2_B m^2_{SUSY}/8
\end{array}
\right]
\eeq
where $m_{SUSY} \sim 300 $ GeV  is of order the slepton  and neutralino masses.

We can match this onto the neutrino mass matrix
for degenerate light neutrinos, with $\theta_{13} = 0$.
Concentrating first on the $\mu \tau$ submatrix, we obtain
\beq
\delta_\mu \simeq \sqrt{\frac{m_1}{m_{SUSY}}} ~~~~~~~
\delta_B \simeq \frac{8 \pi}{g} \sqrt{\frac{m_1}{m_{SUSY}}}
\eeq
and get the large atmospheric  mixing by taking
$m_1$,  the lightest mass of the degenerate neutrinos,
to be $\sqrt{ 4 \pi \Delta m_{atm}^2/g^2 }$ $ \simeq .2$ eV.

The first row has a desirable sign difference between the
$e\mu$ and $e \tau$ entries, and can be adjusted to give
the solar  mass difference and mixing angle by taking
$\lambda \sim .02$ and $\tan \beta \gsim 10$ ($\tan  \beta$
enters via the charged lepton Yukawas). 
%IS THIS CONSISTENT
%WITH RARE PROCESSES?

\subsection*{Acknowledgements}
We thank  Aldo Deandrea, Gino Isidori, and
Alessandro Strumia  for discussions.


\begin{thebibliography}{222222}

\bibitem{DGIS}
G.~D'Ambrosio, G.~F.~Giudice, G.~Isidori and A.~Strumia,
  %``Minimal flavour violation: An effective field theory approach,''
  Nucl.\ Phys.\ B {\bf 645} (2002) 155
  [arXiv:hep-ph/0207036].
  %%CITATION = HEP-PH 0207036;%%

\bibitem{Paolo}
A.~J.~Buras, P.~Gambino, M.~Gorbahn, S.~Jager and L.~Silvestrini,
  %``Universal unitarity triangle and physics beyond the standard model,''
  Phys.\ Lett.\ B {\bf 500} (2001) 161
  [arXiv:hep-ph/0007085].
  %%CITATION = HEP-PH 0007085;%%
C.~Bobeth, M.~Bona, A.~J.~Buras, T.~Ewerth, M.~Pierini, L.~Silvestrini and A.~Weiler,
%``Upper bounds on rare K and B decays from minimal flavor violation,''
Nucl.\ Phys.\ B {\bf 726} (2005) 252
[arXiv:hep-ph/0505110].
%%CITATION = HEP-PH 0505110;%%


%\bibitem{mfv} OTHER MFV?

\bibitem{CGIW}
 V.~Cirigliano, B.~Grinstein, G.~Isidori and M.~B.~Wise,
  %``Minimal flavor violation in the lepton sector,''
  Nucl.\ Phys.\ B {\bf 728} (2005) 121
  [arXiv:hep-ph/0507001].
  %%CITATION = HEP-PH 0507001;%%


\bibitem{numass} 
K.~S.~Babu and C.~N.~Leung,
  %``Classification of effective neutrino mass operators,''
  Nucl.\ Phys.\ B {\bf 619} (2001) 667
  [arXiv:hep-ph/0106054].
  %%CITATION = HEP-PH 0106054;%%
M.~C.~Gonzalez-Garcia and Y.~Nir,
%``Developments in neutrino physics,''
Rev.\ Mod.\ Phys.\  {\bf 75} (2003) 345
[arXiv:hep-ph/0202058].
%%CITATION = HEP-PH 0202058;%%

\bibitem{seesaw}
P.~Minkowski,
%``Mu $\to$ E Gamma At A Rate Of One Out Of 1-Billion Muon Decays?,''
Phys.\ Lett.\ B {\bf 67} (1977) 421;
%%CITATION = PHLTA,B67,421;%%
M. Gell-Mann, P. Ramond and
R. Slansky,  {\em Proceedings of the Supergravity Stony Brook Workshop}, New
York 1979,  eds. P. Van Nieuwenhuizen and D. Freedman; T. Yanagida,  {\em
Proceedinds of the Workshop on Unified Theories and Baryon Number in the
Universe},  Tsukuba, Japan 1979, ed.s A. Sawada and A. Sugamoto;
R. N. Mohapatra, G. Senjanovic,
{\it Phys.Rev.Lett.} {\bf 44} (1980)912.
%

\bibitem{RPV}
 H.P. Nilles and N. Polonsky, Nucl. Phys. {\bf B499} (1997) 33.
 T. Banks, Y. Grossman, E. Nardi and Y. Nir,  
Phys. Rev. {\bf D52} (1995) 5319.
L.~Hall and M.~Suzuki.
 {\em Nucl. Phys.}, B231:419, 1984.
 Y. Grossman and H. Haber, {\it Phys. Rev. Lett.} 
 {\bf 78} (1997) 3438;
{\it   Phys.Rev.} {\bf D59} 093008;  hep-ph/9906310.
 R. Hempfling {\it Nucl.Phys.} {\bf B478} (1996) 3.
 Eung Jin Chun, Sin Kyu Kang,
{\it  Phys.Rev.}  {\bf D61} (2000) 075012,
M. Hirsch, M.A. Diaz, W. Porod, J.C. Romao, J.W.F.
Valle,  hep-ph/0004115. 



\bibitem{Zee}
 A.~Zee,
  %``A Theory Of Lepton Number Violation, Neutrino Majorana Mass, And
  %Oscillation,''
  Phys.\ Lett.\ B {\bf 93} (1980) 389
  [Erratum-ibid.\ B {\bf 95} (1980) 461].
  %%CITATION = PHLTA,B93,389;%%



\bibitem{triplet} G.~B.~Gelmini and M.~Roncadelli,
  %``Left-Handed Neutrino Mass Scale And Spontaneously Broken Lepton Number,''
  Phys.\ Lett.\ B {\bf 99} (1981) 411.
  %%CITATION = PHLTA,B99,411;%% REFS FOR THE MODEL
J.~Schechter and J.~W.~F.~Valle,
  %``Neutrino Decay And Spontaneous Violation Of Lepton Number,''
  Phys.\ Rev.\ D {\bf 25} (1982) 774.
  %%CITATION = PHRVA,D25,774;%% 
A.~Santamaria,
  %``The Hyperchargeless Triplet Majoron Model,''
  Phys.\ Rev.\ D {\bf 39} (1989) 2715.
  %%CITATION = PHRVA,D39,2715;%%
K.~Choi and A.~Santamaria,
  %``17-KeV neutrino in a singlet - triplet majoron model,''
  Phys.\ Lett.\ B {\bf 267} (1991) 504.
  %%CITATION = PHLTA,B267,504;%%
C.~Wetterich,
  %``Neutrino Masses And The Scale Of B-L Violation,''
  Nucl.\ Phys.\ B {\bf 187} (1981) 343.
  %%CITATION = NUPHA,B187,343;%%

\bibitem{anna}
A.~Rossi,
  %``Supersymmetric seesaw without singlet neutrinos: Neutrino masses and
  %lepton-flavour violation,''
  Phys.\ Rev.\ D {\bf 66} (2002) 075003
  [arXiv:hep-ph/0207006].
  %%CITATION = HEP-PH 0207006;%%

%\bibitem{others}



\bibitem{CKM}see {\it e.g.} A.~J.~Buras,
%``Flavour physics and CP violation,''
arXiv:hep-ph/0505175.
%%CITATION = HEP-PH 0505175;%%



\bibitem{Belen}
S.~Antusch, C.~Biggio, E.~Fernandez-Martinez, M.~B.~Gavela and J.~Lopez-Pavon,
%``Unitarity of the leptonic mixing matrix,''
arXiv:hep-ph/0607020.
%%CITATION = HEP-PH 0607020;%%



\bibitem{CG}
 V.~Cirigliano and B.~Grinstein,
  %``Phenomenology of minimal lepton flavor violation,''
  arXiv:hep-ph/0601111.
  %%CITATION = HEP-PH 0601111;%%
V.~Cirigliano, G.~Isidori and V.~Porretti,
%``CP violation and leptogenesis in models with minimal lepton flavour
%violation,''
arXiv:hep-ph/0607068.
%%CITATION = HEP-PH 0607068;%%




\bibitem{9810328}
G.~C.~Branco, M.~N.~Rebelo and J.~I.~Silva-Marcos,
%``Degenerate and quasi degenerate Majorana neutrinos,''
Phys.\ Rev.\ Lett.\  {\bf 82} (1999) 683
[arXiv:hep-ph/9810328].
%%CITATION = HEP-PH 9810328;%%


\bibitem{hisano1}
J.~Hisano, T.~Moroi, K.~Tobe and M.~Yamaguchi,
  %``Lepton-Flavor Violation via Right-Handed Neutrino Yukawa Couplings in
  %Supersymmetric Standard Model,''
  Phys.\ Rev.\ D {\bf 53} (1996) 2442
  [arXiv:hep-ph/9510309].
  %%CITATION = HEP-PH 9510309;%%

\bibitem{LFVss} an incomplete selection of papers
studying LFV in the seesaw (see also \cite{hisano1}): \\
J.~Hisano and D.~Nomura,
  %``Solar and atmospheric neutrino oscillations and 
%lepton flavor violation  in
  %supersymmetric models with the right-handed neutrinos,''
  Phys.\ Rev.\ D {\bf 59} (1999) 116005
  [arXiv:hep-ph/9810479].
  %%CITATION = HEP-PH 9810479;%%
\\
S.~Lavignac, I.~Masina and C.~A.~Savoy,
  %``Large solar angle and seesaw mechanism: A bottom-up perspective,''
  Nucl.\ Phys.\ B {\bf 633} (2002) 139
  [arXiv:hep-ph/0202086].
  %%CITATION = HEP-PH 0202086;%%
\\
J.~A.~Casas and A.~Ibarra,
  %``Oscillating neutrinos and mu $\to$ e, gamma,''
  Nucl.\ Phys.\ B {\bf 618} (2001) 171
  [arXiv:hep-ph/0103065].
  %%CITATION = HEP-PH 0103065;%%

\bibitem{DI}
S.~Davidson and A.~Ibarra,
  %``Determining seesaw parameters from weak scale measurements?,''
  JHEP {\bf 0109} (2001) 013
  [arXiv:hep-ph/0104076].
  %%CITATION = HEP-PH 0104076;%%


\bibitem{DL2}S.~Davidson and M.~Losada,
  %``Basis independent neutrino masses in the R(p) violating MSSM,''
  Phys.\ Rev.\ D {\bf 65} (2002) 075025
  [arXiv:hep-ph/0010325].
  %%CITATION = HEP-PH 0010325;%%

\bibitem{neudata}
B.~Aharmim {\it et al.}  [SNO Collaboration],
%``Electron energy spectra, fluxes, and day-night asymmetries of B-8 solar
%neutrinos from the 391-day salt phase SNO data set,''
Phys.\ Rev.\ C {\bf 72} (2005) 055502
[arXiv:nucl-ex/0502021].
%%CITATION = NUCL-EX 0502021;%%
T.~Araki {\it et al.}  [KamLAND Collaboration],
%``Measurement of neutrino oscillation with KamLAND: Evidence of spectral
%distortion,''
Phys.\ Rev.\ Lett.\  {\bf 94} (2005) 081801
[arXiv:hep-ex/0406035].
%%CITATION = HEP-EX 0406035;%%
J. Nelson,  for the Minos Experiment,  talk at Neutrino'06, Santa Fe. 
 Y.~Ashie {\it et al.}  [Super-Kamiokande Collaboration],
   ``A measurement of atmospheric neutrino oscillation parameters by
  %Super-Kamiokande I,''
  Phys.\ Rev.\ D {\bf 71} (2005) 112005
  [arXiv:hep-ex/0501064].
  %%CITATION = HEP-EX 0501064;%%




\bibitem{PDG}
W.-M. Yao et al., {\it J. Phys.} {\bf  G 33}, 1 (2006)
 
\bibitem{meg}
M.~L.~Brooks {\it et al.}  [MEGA Collaboration],
%``New limit for the family-number non-conserving decay mu+ $\to$ e+ gamma,''
Phys.\ Rev.\ Lett.\  {\bf 83} (1999) 1521
[arXiv:hep-ex/9905013].
%%CITATION = HEP-EX 9905013;%%


\bibitem{meee}
U.~Bellgardt {\it et al.}  [SINDRUM Collaboration],
%``Search For The Decay Mu+ $\to$ E+ E+ E-,''
Nucl.\ Phys.\ B {\bf 299} (1988) 1.
%%CITATION = NUPHA,B299,1;%%


\bibitem{EFT}M.~Pospelov, A.~Ritz and Y.~Santoso,
  %``Flavor and CP violating physics from new supersymmetric thresholds,''
  Phys.\ Rev.\ Lett.\  {\bf 96} (2006) 091801
  [arXiv:hep-ph/0510254].
  %%CITATION = HEP-PH 0510254;%%
E.~J.~Chun,
%``Testing neutrino mass models,''
AIP Conf.\ Proc.\  {\bf 805} (2006) 145
[arXiv:hep-ph/0510318].
%%CITATION = HEP-PH 0510318;%%


\bibitem{Davidson:2000uc}
S.~Davidson and M.~Losada,
%``Neutrino masses in the R(p) violating MSSM,''
JHEP {\bf 0005}, 021 (2000)
[arXiv:hep-ph/0005080].
%%CITATION = HEP-PH 0005080;%%





\end{thebibliography}
\end{document}